**Characterizing the Heterogeneity of the OpenStreetMap Data and Community**

Ding Ma, Mats Sandberg and Bin Jiang

Faculty of Engineering and Sustainable Development
University of Gävle, SE-801 76 Gävle, Sweden
Email: ding.ma|mats.sandberg|bin.jiang@hig.se

*(Draft: October 2014, Revision: February, March 2015)*

**Abstract:** OpenStreetMap (OSM) constitutes an unprecedented, free, geographic information source contributed by millions of individuals, resulting in a database of great volume and heterogeneity. In this study, we characterize the heterogeneity of the entire OSM database and historical archive in the context of big data. We consider all users, geographic elements, and user contributions from an eight-year data archive, at a size of 692 GB. We rely on some nonlinear methods such as power-law statistics and head/tail breaks to uncover and illustrate the underlying scaling properties. All three aspects (users, elements, and contributions) demonstrate striking power laws or heavy-tailed distributions. The heavy-tailed distributions imply that there are far more small elements than large ones, far more inactive users than active ones, and far more lightly edited elements than heavily edited ones. Furthermore, about 500 users in the core group of the OSM are highly networked in terms of collaboration.

## 1. Introduction

Twenty-first century society benefits considerably from, and is increasingly driven by, two forces characterized by the head and the tail of a long-tail distribution (Anderson 2006). For example, while the telephone industry was dominated by national telecoms such as AT&T, we now have services such as Skype. The Encyclopedia Britannica was very popular, but we now have a free, more popular counterpart in Wikipedia. Information was controlled by governments and mass media giants such as CNN, but WikiLeaks or OpenLeaks recently made history by freely sharing information. In the same vein, volunteered geographic informaton (VGI) (Goodchild 2007) emerged as a counterpart to geographic information, which is conventionally collected and maintained by national mapping agencies. As part of user-generated content in the era of Web 2.0, VGI is uniqueby providing geo-referenced location information. OpenStreetMap is the most successful and well-known project of VGI. It attracts significant sustained interest in academia, industry, and government.

In this article, we study all OSM data collected over the past decade, submitted by about 1 million registered users up to February 2013. Previous studies showed that both the data and the user community are very heterogeneous. For example, only a small percentage of users make almost all the contributions, including creation and edits (Neis and Zipf 2012, Mooney and Corcoran 2012a, 2012b). In terms of data concentration and accuracy, the OSM data varies dramatically from urban to rural areas, or from country to country (Neis et al. 2011, Neis and Zielstra 2014). However, these previous studies were conducted mostly at country and city levels. They lack quantitative indicators about heterogeneity or variation. In contrast, we examined all the OSM data and its history to present a holistic picture of OSM based on power-law statistics and the head/tail breaks-induced ht-index. More specifically, we illustrate and quantify the underlying heterogeneity of the OSM elements, the users, and their contributions through a set of quantitative metrics such as α, p value and ht-index.

Power-law statistics is based on the robust maximum-likelihood estimation, which differs from the conventional least-square estimation (Clauset et al. 2009) (see Section 3 for more details). The maximum-likelihood estimation provides two metrics: α (degree of heterogeneity), and p value (goodness of fit). On the other hand, the head/tail breaks (Jiang 2013) is a newly developed classification scheme for data with a heavy-tailed distribution. It also is an efficient, effective visualization tool for big data (Jiang 2015). Head/tail breaks partition the whole around an average size into many small things in the tail being a majority, and a few large ones in the head being a minority. This partition continues recursively for the head (the large things) until the notion of far more small things than large ones is violated. Eventually, the number of times that far more small things recurs is defined as the ht-index (Jiang and Yin 2014) for characterizing complexity or hierarchical levels of the whole.

This paper's contribution is three-fold. We situated the study in the context of big data and extracted the related historical and attributed information from the entire OSM databases and users' historic archive. Based on the extraction, we characterized the heterogeneity of OSM databases and discovered very striking scaling patterns for both users and data. We built up the co-contribution networks over the eight-year timespan of the data and found the underlying nonlinear characteristics of user collaboration networks.

The remainder of the paper is organized as follows: Section 2 presents the OSM history, data, and the working procedure of processing the huge data set. Section 3 briefly introduces the methodology for conducting the scaling analysis, including power-law statistics, detection, and head/tail breaks. Section 4 shows the statistical results of the scaling patterns and other results. Section 5 further discusses this study's implications. Finally, Section 6 draws conclusions and points to future work.

## 2. Data and data processing

Started in July 2004, and motivated by the great success of Wikipedia, OSM aimed to provide free editable maps for the entire world (Bennett 2010). A large number of volunteers relied on GPS receivers to collect trajectory data and transformed it into map data using online editing tools. The mapping processes are time-consuming and tedious. In 2006, Yahoo! donated digital images to the OSM community, so that mapping could be done directly from the images. Later on, OSM obtained free data sets from companies and countries, such as a complete road data set of Netherlands donated by Automotive Navigation Data, and the transformation of a US Census TIGER road data set. Over the past decade, OSM became one of the largest geodata sources and most famous VGI platforms, with around 1.8 million users and billions of geographic elements.

The OSM data is freely accessed on the Internet, with a number of supported formats such as XML and shape files. This study uses the complete, global, OSM data-history dump (http://planet.openstreetmap.org/planet/full-history/). The dump is large, at 692 GB collected from April 9, 2005 to Feb. 5, 2013. It mainly includes, and is structured sequentially by, three basic types of geographical elements of OSM data: Node, way, and relation. Nodes are point features that store the location information of longitude and latitude coordinates. Ways are polylines and polygons that contain a set of ordered nodes. Relation denotes the geographic relationships among the three types of elements. Each element contains a variety of information, such as user and element ID, timestamp of creation or edits, contributing user, version number, and different kinds of tags. The historical information is organized by version numbers with the attribute name *version*, which increases by 1 each time there is a new version of this element.

It is difficult to work with such a big file, since simply running it takes several hours on a state-of-the-art desktop computer. We developed a working procedure (Figure 1) to extract both historical and attribute information for each element of the entire database for further analysis. For the historical information, we collected element ID, timestamp, contributing user ID and version number at each version. Attribute information of each element was with respect to the latest version. For each node element, we extracted its coordinate pair (latitude and longitude), and for each way and relation element, we collected their member IDs. The whole process took three days on an eight-core, 3.4-GHz CPU, 32-GB memory desktop. The extraction was organized as a big table and formatted as a .txt file with a size of about 150 GB, including approximate 2.1 billion elements consisting of 1.9 billion nodes, 0.2 billion ways and 2 million relations. For further analysis, we calculated the number of users, edits, and sizes for each element and their spatial distribution at the country level (see detailed description in Section 4). The extraction as a shrunk version of data greatly improves efficiency, as it only takes half an hour to traverse the file and less than one second to return query results by using a binary search based on sorted element ID.

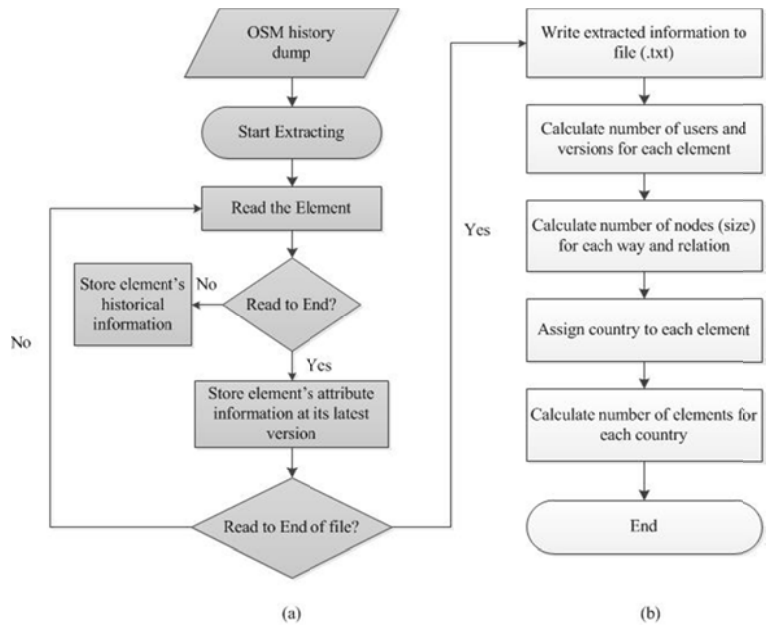

Figure 1: Flow chart for the data processing to extract essential information for further analysis
Note that (a) is data processing and (b) is results

## 3. Methodology

This study illustrates a set of adopted nonlinear methods, including power-law detection based on maximum-likelihood estimation (Clauset et al. 2009) and head/tail breaks as a classification scheme for data with a heavy-tailed distribution (Jiang 2013). We rely on these two methods for scaling analysis because power-law detection is probably the most robust, reliable method for power-law estimation, and head/tail breaks are both a classification scheme and visualization tool (Jiang 2015a). These two methods greatly complement each other to uncover and visualize the underlying scaling properties of OSM data. More specifically, power-law detection is concerned with how a data set fits power laws better than alternatives such as lognormal, exponents, and their variants. Head/tail breaks reveals the inherent hierarchical levels, or the head/tail breaks-induced ht-index (Jiang and Yin 2013). More importantly, the head/tail breaks can efficiently, effectively filter out redundant data, as a powerful visualization tool for big data.

### 3.1 Power law detection

Data bearing the scaling property follows a power-law distribution, which means that the frequency of each value is inversely proportional to the power of its rank. In other words, the data contains far more small values yet very few large values. The most famous example of power law is found in word occurrences, city sizes, and wealth distributions (Zipf 1949). Generally, the power law is denoted as:

$$y = kx^{-\alpha} \qquad [1]$$

in which $k$ is a constant, and $\alpha$ is the power law exponent.

The simplest way to detect the power law is to take logarithm scale on both axes to see if the distribution curve is a straight line, based on:

$$\ln(y) = -\alpha \ln(x) + \ln(k) \qquad [2]$$

However, this method suffers from the messy tail at the very end of the distribution. Clauset et al. (2009) introduced a rigorous statistical test based on maximum likelihood and the Kolmogorov-Smirnov (KS) test for power-law detection. There are two parameters: An estimated exponent $\alpha$ and the index of a goodness-of-fit $p$. They are used as indices for power-law fit and the goodness of the fit. This method has been widely used and proven robust for detecting the power-law distributions with a wide range of complex systems (Marta et al. 2008, Jiang et al. 2009, Jiang and Jia 2011).

Simply put, the estimated exponent $\alpha$ shapes the power-law distribution and the acceptance range is from 1 to 3, given by:

$$\alpha = 1 + n \left[ \sum_{i=0}^{n} \ln \frac{x_1}{x_{min}} \right]^{-1} \qquad [3]$$

in which α denotes the estimated exponent, and $x_{min}$ is the smallest value above which the power-law fit is held.

We adopted a modified KS test to assess how data fits a power-law distribution (goodness of fit). It is based on the idea of the maximum distance ($D$) between the cumulative density functions (CDF) of the data and the fitted model:

$$D = \max_{x \geq x_{min}} |f(x) - g(x)| \qquad [4]$$

in which $f(x)$ is the CDF of the data for the observations with a value of at least $x_{min}$, and $g(x)$ is the CDF for the power-law model that best fits the data in which $x \geq x_{min}$.

Usually, 1,000 synthetic data sets are then generated with the fitted model $g(x)$, which contains data whose values above $x_{min}$ perfectly follow a power-law distribution. Conversely, values below $x_{min}$ are not power-law distributed. The maximum difference $D$ is re-calculated between the fitted model and each synthetic dataset. The goodness-of-fit index $p$ is denoted as a fraction of the number of $D_i$ whose values are greater than $D$ to 1,000. The higher the p value, the better fit with the power law. The closer the p-value gets to 1, the more the data is accepted for a power-law distribution. The acceptable threshold for goodness of fit is 0.05.

Power-law detection is probably the toughest statistical estimation to differentiate power laws from other alternatives, such as lognormal, exponential, and other variants. In contrast to the rigorous power-law detection, the head/tail breaks provides a simple solution to reveal the underlying scaling. It applies for all kinds of heavy-tailed distributions, as long as the scaling pattern of far more small things than large ones recurs multiple times.

### 3.2 Head/tail breaks

The head/tail breaks is basically originated from the main characteristic of heavy-tailed distributions. Given data with a heavy-tailed distribution, the arithmetic mean, or average, can split all the data values into two unbalanced parts: A minority of big values above the mean, called the *head*; and a majority of small values below the mean, called the *tail*. This process recursively continues for the head until the notion of far more small values than large ones is violated; see the following recursive function namely head/tail breaks. The percentage of splitting up data into the head and tail is set at 40 percent. This implies that the tail percentage is 60 percent. The number of times the data can be split + 1 is the ht-index (Jiang and Yin 2014). It captures how many times the scaling pattern of far more small things than large ones recurs in the data. It quantifies the scaling characteristic of the data. The higher the ht-index, the more hierarchical levels in the data.

```
Recursive function Head/tail Breaks:
```

```
    Break the input data (around mean or average) into the head and the
        tail;
    // the head for data values above the mean
    // the tail for data values below the mean
    while (head <= 40%):
        Head/tail Breaks(head);
End Function
```

Some data in this study, such as 2 billion elements, were too big to detect the power laws. In this regard, head/tail breaks provide a nice solution. Instead of taking all the elements, we took the head part for power-law detection. If the head part was still too big, we took the next head part, until the head part was small enough for power-law detection. The reason why we recursively take the head is simply because the head is self-similar to the whole data set. This is also the fundamental argument for the head/tail breaks as an efficient, effective visualization tool for big data (Jiang 2015a). Therefore, power-law detection and head/tail breaks complement each other and provide powerful tools for revealing the underlying scaling or heterogeneity of the OSM data.

## 4. Scaling properties of the OSM data

This section presents the results of the scaling analysis on a variety of features based on three aspects in the context of big data (1 million users, 2.1 billion elements, and 2.7 billion contributions) (Figure 2). These three parts constitute an interconnected picture of the OSM data and community. The users contribute to the elements, leading to a great increase in both element volume and complexity, and the user community. Through the contributions, users formed an interconnected collaboration network. The scaling analysis based on power-law detection and head/tail breaks was applied to these three aspects to examine to what extent the scaling pattern of far more small things than large ones was true for the OSM data.

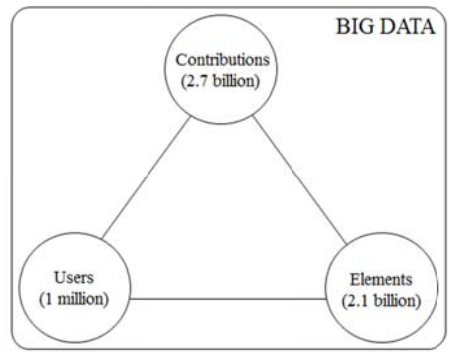

Figure 2: Three aspects of the study in the context of big data

### 4.1 On users and elements

We first investigated users based on their number of contributions. The investigation was based on how many unique element IDs can one user contribute to. These contributions include both creating and editing. A total of 268,227 users made contributions. The number of each user's contributions exhibited a power-law distribution, with an accepted $\alpha$ of 2.24 and p value of 0.26 (Figure 3a). By applying head/tail breaks, we derived the scaling hierarchy of these numbers, indicated by the ht-index of 7 and very low percentages for each head (< 30 percent). This means the scaling pattern recurred six times in this data (Table 1). This apparent scaling pattern indicates that only a very small number of users contributed the majority of OSM elements. In other words, there were far more inactive users than active ones.

Table 1: Head/tail breaks statistics for user contributions
(Note: # = number, % = percentage)

| #Sum | #head | %head | # tail | %tail | mean |
|---|---|---|---|---|---|
| 268,227 | 13,241 | 4% | 254,986 | 96% | 10,232 |
| 13,241 | 1,825 | 13% | 11,416 | 87% | 199,020 |
| 1,825 | 307 | 16% | 1,518 | 84% | 1,164,797 |
| 307 | 48 | 15% | 259 | 85% | 4,751,085 |
| 48 | 8 | 16% | 40 | 84% | 18,843,785 |
| 8 | 2 | 25% | 6 | 75% | 69,899,060 |

Second, we looked at different attributes of elements. Each element was characterized by the number of users, edits, and sizes. Specifically, the number of users for each element indicated how many users contributed to it, given by the number of unique user IDs for each element. Contributions included both creating and editing. The number of edits was directly obtained by the maximum version number of each element, since it equaled the maximum version number minus 1. The number of size refers to how many unique node IDs it contained. The size of each node element was always 1. The size of each way element equaled the number of its unique comprising points. The size of each relation element was determined by the number of unique points its member contained: node, way, or relation (these three parts do not always exist simultaneously in one relation). Because some relation elements can have other relation element(s) as its member(s), it is difficult to calculate those relation elements' sizes when they mutually contain each other as their member(s). There were 4,356 relation elements excluded because of such complicated structures. Considering the elements of 2.1 billion elements studied, we believe that the excluded relations would not affect much our results.

Next, we applied head/tail breaks to the above three aspects. All three derived ht-indices were very high (> 10), and most of the head percentages were small (< 30 percent; see detailed results in the appendix). It indicated that there were far more small elements than large ones. Power-law detection was further applied to the data for the top hierarchical levels of each category (Table 2). The filtered data was the proxy of the whole, since the scaling pattern remains at each level. Only the number of element sizes passed the power-law test (Figure 4d). The number of element users and edits was still heavy-tail distributed. Figures 3b and 3c show that each plot was close to a straight line at logarithm scales. We think that the entire data set of three aspects possess a strong scaling property. We also examined the evolution of data on a yearly basis and found that heterogeneity was no different from the data as a whole. In other words, the data for the previous years are all heavy-tail distributed, but vary with different ht-indices.

Table 2: Summarized statistics on OSM elements for top hierarchies in three categories

| | # element | max | min | α | p |
|---|---|---|---|---|---|
| User | 745,943 | 197 | 7 | 4.95 | 0 |
| Edit | 548,914 | 3,084 | 31 | 3.39 | 0.006 |
| Size | 479,004 | 5,118,276 | 564 | 2.37 | 0.13 |

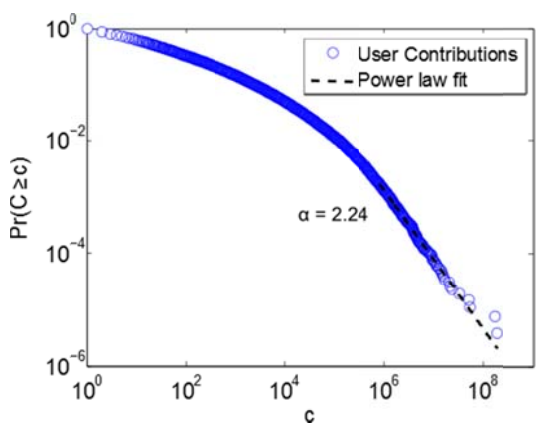
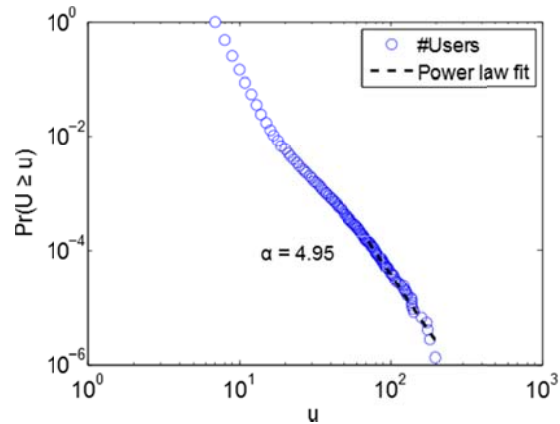

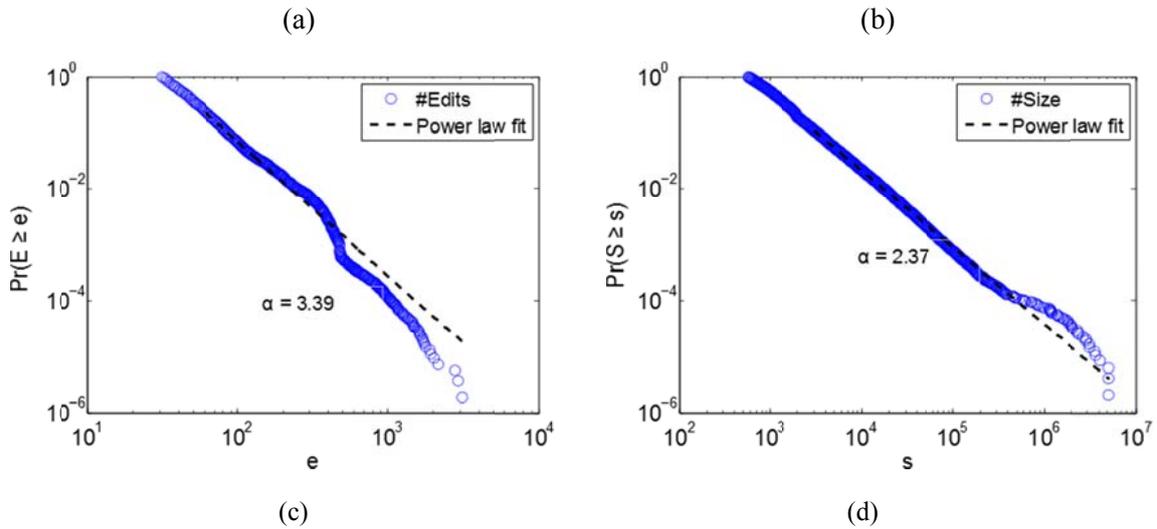

Figure 3: Power-law distributions of user contributions: (a) number of users, (b) number of edits, and (c) number of sizes (d) of each element. The data for (b), (c), and (d) are selected from the top hierarchies of all elements. (b) and (c) are not power-law distributed because both α values are larger than 3, but they are heavy-tailed, illustrated by the high ht-index shown in the Appendix.

We further inspected the spatial distribution of the elements, or how many of the elements were located in each country. We computed and assigned to each country the number of elements and aggregated attribute values of each aspect. The data of all three aspects was very power-law distributed (Table 3). This indicated that there were far more small countries than large ones in terms of elements, users, and contributions. It further implies the great variation of quality and completeness of OSM databases from country to country through different elements concentrations. The cartogram shows the resulting country sizes (Figure 4). The top five countries were: US, France, Canada, Germany, and Russia. These countries were also the top five in terms of aggregated numbers of users, edits, and sizes, but with a slightly different ranking (Canada and Germany switched positions).

Table 3: Summarized statistics of elements at country level

|          | max         | min   | α    | p    |
|----------|-------------|-------|------|------|
| #Element | 401,137,304 | 836   | 1.74 | 0.87 |
| #User    | 598,175,441 | 951   | 1.74 | 0.82 |
| #Edit    | 636,597,363 | 969   | 1.73 | 0.71 |
| #Size    | 898,145,600 | 1,408 | 1.69 | 0.67 |

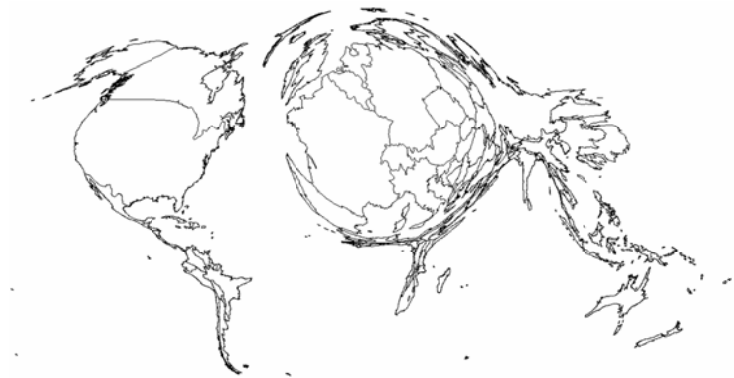

Figure 4: The cartogram showing the spatial distribution of global OSM elements at country level

## 4.2 On a co-contribution network

Having examined users and elements, we subsequently studied the scaling pattern in the collaboration network of the OSM users. The social relationship utilized in this research is a co-contribution relationship, which was established in the OSM data archive when more than one user contributed to the same element. In other words, we considered that a user had such relationship with others if they either created or edited the same element. This approach is different from the one Mooney and Corcoran (2012a, 2012b) defined. They only considered the edit interaction and also the sequence of edits. In this regard, we constructed a co-contribution network, rather than a co-edit network. As Figure 5a shows, assuming that users 1, 3, and 4 make contributions to Element b, there are co-contribution relationships between every two of them, so that the resulting network (Figure 5b) can be obtained. This paper only considers the binary network, which consists of undirected, unweighted edges.

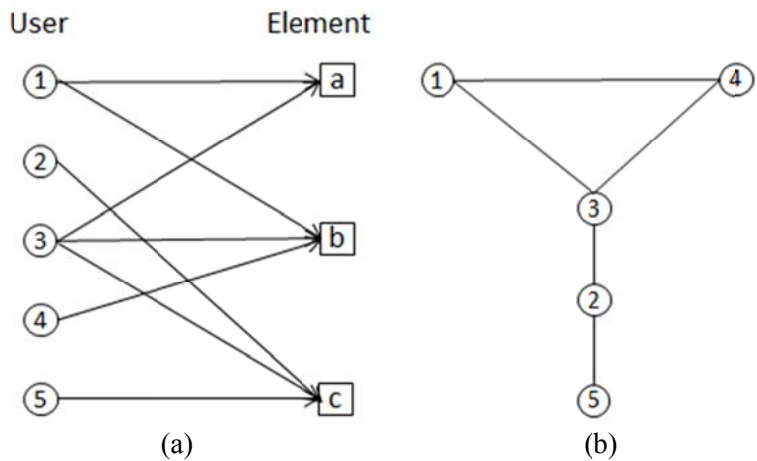

(a)          (b)

Figure 5: Illustration of co-contribution relationship
Note: Users' contributions to elements are represented as a bi-partite graph (a), which is transformed into a co-contribution network (b)

Table 4: Head/tail breaks statistics for node degree of co-contribution network in 2013

| #nodes | #head | %head | #tail | %tail | mean |
| --- | --- | --- | --- | --- | --- |
| 248,070 | 33,504 | 13% | 214,566 | 87% | 51.97 |
| 33,504 | 7,267 | 21% | 26,237 | 79% | 322.08 |
| 7,267 | 1,820 | 25% | 5,447 | 75% | 1037.53 |
| 1,820 | 477 | 26% | 1,343 | 74% | 2486.16 |
| 477 | 137 | 28% | 340 | 72% | 5181.7 |
| 137 | 40 | 29% | 97 | 71% | 9474.48 |
| 40 | 11 | 27% | 29 | 73% | 16102.82 |
| 11 | 3 | 27% | 8 | 73% | 26460.73 |
| 3 | 1 | 33% | 2 | 67% | 47980.33 |

Following the rule of co-contribution relationship, we built the network based on the entire history of all the elements to better illustrate engagement in the OSM community (Hristova et al. 2012). The resulting social graph consists of 248,070 nodes and 6,446,086 edges. The node degree of this network was power-law distributed and had an ht-index of 10 (Table 4), indicating that the network was extremely scale-free. Figure 6 shows the filtered network, comprising 477 nodes of the top five hierarchies as representative of the entire network, from which the underlying scaling pattern is clearly uncovered. We further examined the networks of previous years from 2005 to see if the scaling pattern persisted throughout the evolution of the OSM community in terms of contributions. Except for no such network between 2005 and 2006, the evolution of co-contribution networks had nonlinear growth of both nodes and edges from 2007 onward and became increasingly scaled. This is indicated by the power-law fitting metrics and an overall increasingly large ht-index of each year (Table 5).

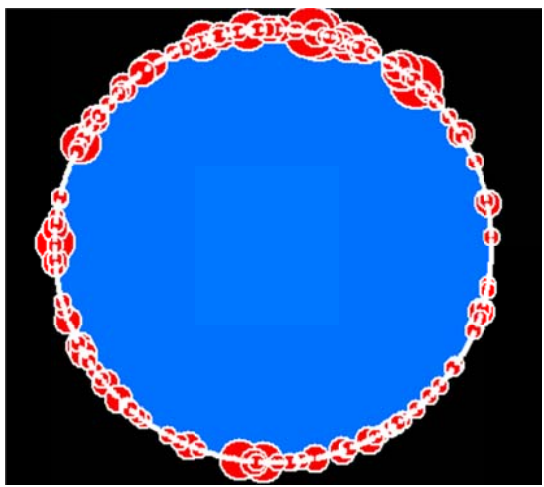

Figure 6: The co-contribution network for the top five hierarchical levels involving 477 nodes and 80,957 edges
Note: The scaling hierarchy of far more small nodes than larger ones is indicated by the size of red dots

Table 5: Scaling analysis results of co-contribution networks from 2007 to 2013

|           | 2007   | 2008    | 2009      | 2010      | 2011      | 2012      | 2013      |
|-----------|--------|---------|-----------|-----------|-----------|-----------|-----------|
| # of nodes | 3,856  | 25,133  | 60,231    | 101,364   | 159,747   | 240,119   | 248,070   |
| # of edges | 29,701 | 418,077 | 1,306,154 | 2,415,319 | 3,954,826 | 6,192,510 | 6,446,086 |
| max-degree | 802    | 5,449   | 15,052    | 30,816    | 51,501    | 65,190    | 65,876    |
| ht-index   | 8      | 7       | 9         | 9         | 9         | 10        | 10        |
| A          | 2.8    | 2.68    | 2.57      | 2.59      | 2.64      | 2.53      | 2.91      |
| P          | 0.46   | 0.65    | 0.18      | 0.14      | 0.06      | 0         | 0.2       |

We also developed some insights into the OSM community in terms of user collaboration from the derived co-contribution network. Compared to the collaboration network of English-language Wikipedia (Laniado and Tasso 2011, Voss 2005, Hirth et al. 2012), it has the same scaling pattern of far more inactive users than active ones. In addition, the network also has some high-density concentrations, especially among highly active users. Specifically, each user collaborates, on average, with around 52 other users in the whole network. The high-degree users collaborate with almost every other user. We further selected two global location-based social networks (Gowalla and Brightkite) for comparison (data available at http://snap.stanford.edu/data/index.html#canets) and found that the co-contribution network was much denser than either social network regarding both the whole and sampled (nodes of top hierarchies) network.

## 5. Further discussions on the study
From the results presented in Section 4, we found a great heterogeneity of OSM data and community, from elements, to user contributions, to the annual co-contributions networks. All are well-characterized by the striking scaling patterns, which are indicated by some metrics of power-law statistics and head/tail breaks. This section further discusses some implications of the results and the study in general.

This study processed and analyzed the entire OSM data and community archive from a holistic perspective, including elements, users, and collaboration networks evolving over the past decade. More than hundreds of gigabytes of the data were processed and computed to develop new insights into the data and community. The findings of this study are in line with previous research on users and elements, in that that a minority of users/elements accounts for a majority of contributions/edits. The major difference between our work and previous studies is that we conducted an in-depth quantitative analysis on all users and elements at the global scale. This enabled us to see something that was not illustrated in previous works. To our best knowledge, the scaling patterns have never been examined for the OSM data set at such a massive level. In this connection, we believe that this study can be extended to other user-generated content such as Wikipedia (Voss 2005).

This paper applies the scaling analysis to characterize the heterogeneity of the global OSM database. Apart from examining the power-law statistics for detecting scaling patterns, other heavy-tailed distributions were observed and measured by the ht-index. It is widely known that data from real-world phenomena is very likely to be heavy-tail distributed, as is the case with the OSM data. The data naturally evolves and accumulates from individuals from the bottom up, rather than imposed by authorities from the top down. As a result, the data of all aspects generally follows power laws or heavy-tailed distributions. Therefore, conventional linear methods such as Gaussian statistics show some inadequacies in characterizing this kind of heterogeneity. There is no typical mean or scale to characterize the heterogeneity. Instead, all scaling characterizes the diversity or heterogeneity. Our study argues that, in the big-data era, geospatial analysis requires a new way of thinking, or Paretian thinking (Jiang 2015b), to better understanding geographic forms and processes.

Big data, due to its diversity and heterogeneity, is likely to demonstrate the scaling pattern of far more small things than large ones. The large and small things constitute the head and tail, respectively, of a long-tailed distribution. Interestingly, the scaling pattern recurs multiple times, which implies that the things in the head recursively demonstrate the scaling pattern of far more small things than large ones. This recurring scaling pattern is what underlies the new classification scheme called head/tail breaks (Jiang 2013). The head/tail breaks divides things around an average into a few large things in the head and many small things in the tail, and continue recursively for the dividing process for the head until the notion of far more small things than large ones is violated. The head/tail breaks can efficiently, effectively filter out data that is too big to handle by conventional means. This filtering function is also what underlies the visualization function of the head/tail breaks (Jiang 2015a). We believe that the head/tail thinking behind the head/tail breaks is very promising for big data and its analytics.

## 6. Conclusion

OSM data is essentially very heterogeneous, either at the local or global scale. This is because geographic space, or the earth's surface, is very heterogeneous with no average location on the earth's surface. In this paper, we studied the entire OSM database and found that this heterogeneity can be fairly illustrated and measured from elements, users, and their collaborations. For the users, both their contributions and the degree of the co-contribution networks exhibit a clear power-law distribution, which means that there are far more inactive users than active ones. There are also far more small elements than large ones, since their attribute values throughout three categories (number of users, edits, and sizes) are heavy-tail distributed. In addition, the elements assigned to individual countries demonstrate a striking power law. Such a pattern also remains at the country level concerning the spatial distribution of all elements. The head/tail breaks can analyze and visualize the big data in capturing the underlying scaling hierarchies and complement the mathematical power-law detection. To summarize, the scaling property is clearly shown with the OSM data and can well-characterize this great heterogeneity through power law fitting the metrics and underlying the scaling hierarchical levels.

The study was conducted from the big-data perspective, which focuses on the entire database and data-intensive computing (Hey et al. 2009). Therefore, we created a comprehensive image of the heterogeneity of the OSM data and obtained a valuable database with respect to the historical and attributable information of all elements at a certain time point. Interested researchers are always welcome to contact us for further detailed information on the data processing. As for future work, two things should be done. The first is to take the tag information of each element into account and conduct a scaling analysis on them. The second is to study the nonlinear dynamics of both spatial and attributable information of each element at different temporal granularities (such as year, month, and week) to find the underlying mechanism of the evolution of both OSM-community and user-mapping activities.

**Acknowledgement**:
Bin Jiang's work is partially supported by special fund of Key Laboratory of Eco Planning & Green Building, Ministry of Education (Tsinghua University), China

**Appendix: The head/tail breaks statistics for users, edits, sizes**

To supplement the description of the results presented in Section 4.1, this appendix contains the detailed statistics on the head/tail breaks process for the three aspects: users, edits, and sizes. As we can see, all the data have more than 12 hierarchical levels, shown in the level column, and the mean head percentages of all three aspects are less than 30%, which is far less than the default threshold of 40%. Note that for the results of each element size (Table A3), there are 4,356 elements excluded from the calculation, therefore the number of elements is 2,138,154,220 – 4,356 = 2,138,149,864.

Table A1: Head/tail breaks statistics for number of users of each element

| Levels | # Elements | # in head | # in tail | head % | tail % | Mean(user) |
|---|---|---|---|---|---|---|
| Source | 2,138,154,220 | 460,660,386 | 1,739,391,359 | 21% | 79% | 1 |
| Level 1 | 460,660,386 | 63,754,888 | 396,905,498 | 14% | 86% | 2 |
| Level 2 | 63,754,888 | 13,945,213 | 49,809,675 | 22% | 78% | 3 |
| Level 3 | 13,945,213 | 4,423,467 | 9,521,746 | 32% | 68% | 5 |
| Level 4 | 4,423,467 | 1,694,469 | 2,728,998 | 38% | 62% | 6 |
| Level 5 | 1,694,469 | 745,943 | 948,526 | 44% | 56% | 7 |
| Level 6 | 745,943 | 189,004 | 556,939 | 25% | 75% | 8 |
| Level 7 | 189,004 | 64,169 | 124,835 | 34% | 66% | 11 |
| Level 8 | 64,169 | 18,277 | 45,892 | 28% | 72% | 14 |
| Level 9 | 18,277 | 5,230 | 13,047 | 29% | 71% | 19 |
| Level 10 | 5,230 | 1,486 | 3,744 | 28% | 72% | 28 |
| Level 11 | 1,486 | 481 | 1,005 | 32% | 68% | 43 |
| Level 12 | 481 | 166 | 315 | 35% | 65% | 63 |
| Level 13 | 166 | 56 | 110 | 34% | 66% | 85 |
| Level 14 | 56 | 19 | 37 | 34% | 66% | 112 |
| Level 15 | 19 | 6 | 13 | 32% | 68% | 144 |

Table A2: Head/tail breaks statistics for number of edits of each element

| Levels | # Elements | # in head | # in tail | head % | tail % | Mean(edit) |
|---|---|---|---|---|---|---|
| Source | 2,138,154,220 | 649,802,777 | 1,550,248,968 | 30% | 70% | 1 |
| Level 1 | 649,802,777 | 129,015,893 | 520,786,884 | 20% | 80% | 2 |
| Level 2 | 129,015,893 | 29,598,177 | 99,417,716 | 23% | 77% | 4 |
| Level 3 | 29,598,177 | 7,795,319 | 21,802,858 | 26% | 74% | 9 |
| Level 4 | 7,795,319 | 1,999,354 | 5,795,965 | 26% | 74% | 16 |
| Level 5 | 1,999,354 | 548,914 | 1,450,440 | 27% | 73% | 31 |
| Level 6 | 548,914 | 158,071 | 390,843 | 29% | 71% | 56 |
| Level 7 | 158,071 | 42,272 | 115,799 | 27% | 73% | 95 |
| Level 8 | 42,272 | 12,769 | 29,503 | 30% | 70% | 166 |
| Level 9 | 12,769 | 4,740 | 8,029 | 37% | 63% | 275 |
| Level 10 | 4,740 | 1,646 | 3,094 | 35% | 65% | 391 |
| Level 11 | 1,646 | 285 | 1,361 | 17% | 83% | 507 |
| Level 12 | 285 | 102 | 183 | 36% | 64% | 850 |
| Level 13 | 102 | 34 | 68 | 33% | 67% | 1,225 |
| Level 14 | 34 | 12 | 22 | 35% | 65% | 1,669 |
| Level 15 | 12 | 4 | 8 | 33% | 67% | 2,113 |

Table A3: Head/tail breaks statistics for each element size

| Levels | # Elements | # in head | # in tail | head % | tail % | Mean(size) |
|---|---|---|---|---|---|---|
| Source | 2,138,149,864 | 166,538,593 | 2,033,513,152 | 8% | 92% | 3 |
| Level 1 | 166,538,593 | 21,021,688 | 145,516,905 | 13% | 87% | 20 |
| Level 2 | 21,021,688 | 280,1262 | 18,220,426 | 13% | 87% | 110 |
| Level 3 | 2,801,262 | 479,004 | 2,322,258 | 17% | 83% | 564 |
| Level 4 | 479,004 | 78343 | 400,661 | 16% | 84% | 2,240 |
| Level 5 | 78,343 | 13,569 | 64,774 | 17% | 83% | 8,258 |
| Level 6 | 13,569 | 2,215 | 11,354 | 16% | 84% | 29,282 |
| Level 7 | 2,215 | 331 | 1,884 | 15% | 85% | 107,770 |
| Level 8 | 331 | 60 | 271 | 18% | 82% | 440,378 |
| Level 9 | 60 | 22 | 38 | 37% | 63% | 1,618,479 |
| Level 10 | 22 | 8 | 14 | 36% | 64% | 2,895,564 |
| Level 11 | 8 | 3 | 5 | 38% | 62% | 4,116,527 |
| Level 12 | 3 | 1 | 2 | 33% | 67% | 5,069,421 |